\documentclass[12pt,draftclsnofoot,onecolumn]{IEEEtran}

\usepackage{enumitem}
\usepackage{amsmath,graphics,amssymb,epsfig,subfigure,color,cite}
\usepackage{array}
\usepackage{multirow}
\usepackage{enumerate}
\usepackage{algorithm2e}
\usepackage{algpseudocode}
\usepackage{algorithmicx}
\usepackage{algcompatible}
\usepackage{bm}
\usepackage{float}
\usepackage{makecell}
\usepackage{comment}

\newtheorem{thm}{Theorem}

\newtheorem{cor}{Corollary}

\floatname{algorithm}{Procedure}

\begin{document}
		
\title{\huge On the Performance of AoA-Based Physical-Layer Authentication Under Multi-Source Impersonation Attacks}
\author{Jeongbin Kim, \IEEEmembership{Graduate Student Member,~IEEE}, Moonsik Min, \IEEEmembership{Member,~IEEE}, and Yo-Seb Jeon, \IEEEmembership{Member,~IEEE}
\thanks{Jeongbin Kim and Yo-Seb Jeon are with the Department of Electrical Engineering, POSTECH, Pohang, Gyeongbuk 37673, South Korea (e-mail: jb2067@postech.ac.kr; yoseb.jeon@postech.ac.kr).}
\thanks{Moonsik Min is with the School of Electronic and Electrical Engineering, Kyungpook National University, Daegu 41566, South Korea (e-mail: msmin@knu.ac.kr).}
}

\maketitle
\makeatletter
\def\blfootnote{\gdef\@thefnmark{}\@footnotetext}
\makeatother

\begin{abstract}
This paper investigates cooperative impersonation attacks against angle-of-arrival (AoA)-based authentication for a wireless communication receiver equipped with a digital uniform linear array. 
Specifically, we characterize the vulnerability and robustness of this authentication system in relation to adversarial angular diversity and the relative numbers of adversarial sources and receive antennas.
To this end, we first derive the optimal precoding vector for multiple spatially distributed adversarial sources by minimizing the mean squared error between legitimate and adversarial received signals. 
By means of the rank property of Vandermonde matrices, we then establish necessary and sufficient conditions under which the adversarial sources can reproduce the legitimate transmitter's spatial signature. 
With fewer adversarial sources than receive antennas, perfect reproduction is not possible without at least one adversarial source satisfying a specific AoA condition. In the other case, increasing the number of receive antennas can reduce impersonation success probabilities when none of the adversarial sources has an AoA satisfying this condition.
Numerical results validate the theoretical analysis and evaluate the security performance of AoA-based authentication.
\end{abstract}


\section{Introduction}\label{Sec:Intro}
Security threats in wireless communication networks extend beyond unauthorized interception to attacks in which adversaries attempt to deceive legitimate receivers. Over the past few decades, cryptographic encryption and decryption, which underpin upper-layer security protocols, have played a fundamental role in information security. However, conventional cryptographic techniques still face the inherent challenge of high computational and key management overhead \cite{wang2018survey}, 
placing a considerable burden on resource-constrained devices such as Internet of Things devices and unmanned aerial vehicles. As an alternative approach, physical-layer security has attracted considerable attention for its potential to achieve confidentiality and authentication by incorporating the physical properties of wireless channels into security mechanisms \cite{liu2016physical}.

One important application of this paradigm is physical-layer authentication, which enables transmitter verification based on received signal characteristics without relying on conventional cryptographic operations. A variety of physical-layer features have been explored to distinguish legitimate transmitters from malicious adversaries, including received signal strength (RSS) \cite{faria2006detecting}, channel impulse response (CIR) \cite{liu2011robust}, and angle-of-arrival (AoA) \cite{pham2026leveraging}. 
However, the combination of AoA with RSS and CIR fails to provide security gains beyond those attainable with AoA-based authentication alone \cite{pham2026leveraging}. This suggests that RSS and CIR may not be sufficiently reliable authentication features and provides strong motivation for investigating the robustness of AoA-based authentication against impersonation attacks.

There has been extensive prior work on AoA-based authentication under different antenna architectures and impersonation attack models. The authors in \cite{srinivasan2024aoa} demonstrated that an adversary can forge the legitimate transmitter's AoA signature to deceive an authenticator equipped with an analog receive array, even from a different direction. Similarly, in \cite{pourafzal2025cooperative}, the feasibility of cooperative impersonation jamming against hybrid beamforming systems was characterized in terms of the number of jammers and receiver radio frequency chains. More recent studies investigated impersonation attacks against digital uniform linear arrays (ULAs) by analyzing optimal precoding design \cite{pham2026leveraging} and authentication error probabilities \cite{skaperas2026robustness}, while extending the analysis to near-field environments \cite{hassani2026near}. Despite these early research efforts, further investigation is needed to understand how multiple spatially distributed adversarial sources can jointly reproduce the legitimate transmitter's signal as received by a digital ULA. Moreover, it remains unclear whether the number of adversarial sources relative to that of receive antennas affects the feasibility of successful impersonation in AoA-based authentication with a digital ULA.

To fill this research gap, we take the analysis a step further by clarifying the conditions that enable cooperative impersonation against an AoA-based authenticator with a digital ULA, as considered in \cite{pham2026leveraging}, \cite{skaperas2026robustness}.
In this setup, multiple spatially distributed adversarial sources jointly attempt to forge the legitimate transmitter's identity. 
We first derive the optimal precoding design for the adversarial sources that minimizes the mean squared error (MSE) between the signals received at the authenticator from the legitimate transmitter and adversarial sources. Based on the derived precoding, we theoretically characterize the conditions under which the adversarial sources can successfully impersonate the legitimate transmitter for both cases where the number of adversarial sources is smaller than or at least equal to the number of receive antennas. Through numerical results, we validate the theoretical analysis by evaluating the minimum achievable MSE and investigate the impersonation success probability with AoAs estimated by the multiple signal classification (MUSIC) algorithm \cite{schmidt1986multiple}. This study provides deeper insights into the vulnerability and robustness of AoA-based authentication by identifying the conditions that permit or prevent perfect spatial signature reproduction. 


\textit{Notations:} 
For a matrix $\mathbf{A}$, $\operatorname{rank}(\mathbf{A})$, $\mathbf{A}^{\dagger}$, and $\mathcal{R}(\mathbf{A})$ denote its rank, Moore-Penrose pseudoinverse, and column space, respectively. The operators $(\cdot)^{\top}$, $|\cdot|$, $\|\cdot\|$, and $\mathbb{E}[\cdot]$ represent the transpose, absolute value, Euclidean norm, and expectation, respectively. 
The $M$-dimensional zero vector and the $M \times M$ identity matrix are denoted by $\mathbf{0}_M$ and $\mathbf{I}_M$, respectively. 
The notation $\mathcal{CN}(\boldsymbol{\mu},\boldsymbol{\Sigma})$ denotes the complex Gaussian distribution with mean vector $\boldsymbol{\mu}$ and covariance matrix $\boldsymbol{\Sigma}$. The uniform distribution over $[a,b]$ is represented by $\mathcal{U}(a,b)$.

\section{System Model}\label{Sec:System}
\begin{figure}[t]
\centering 
    {\epsfig{file=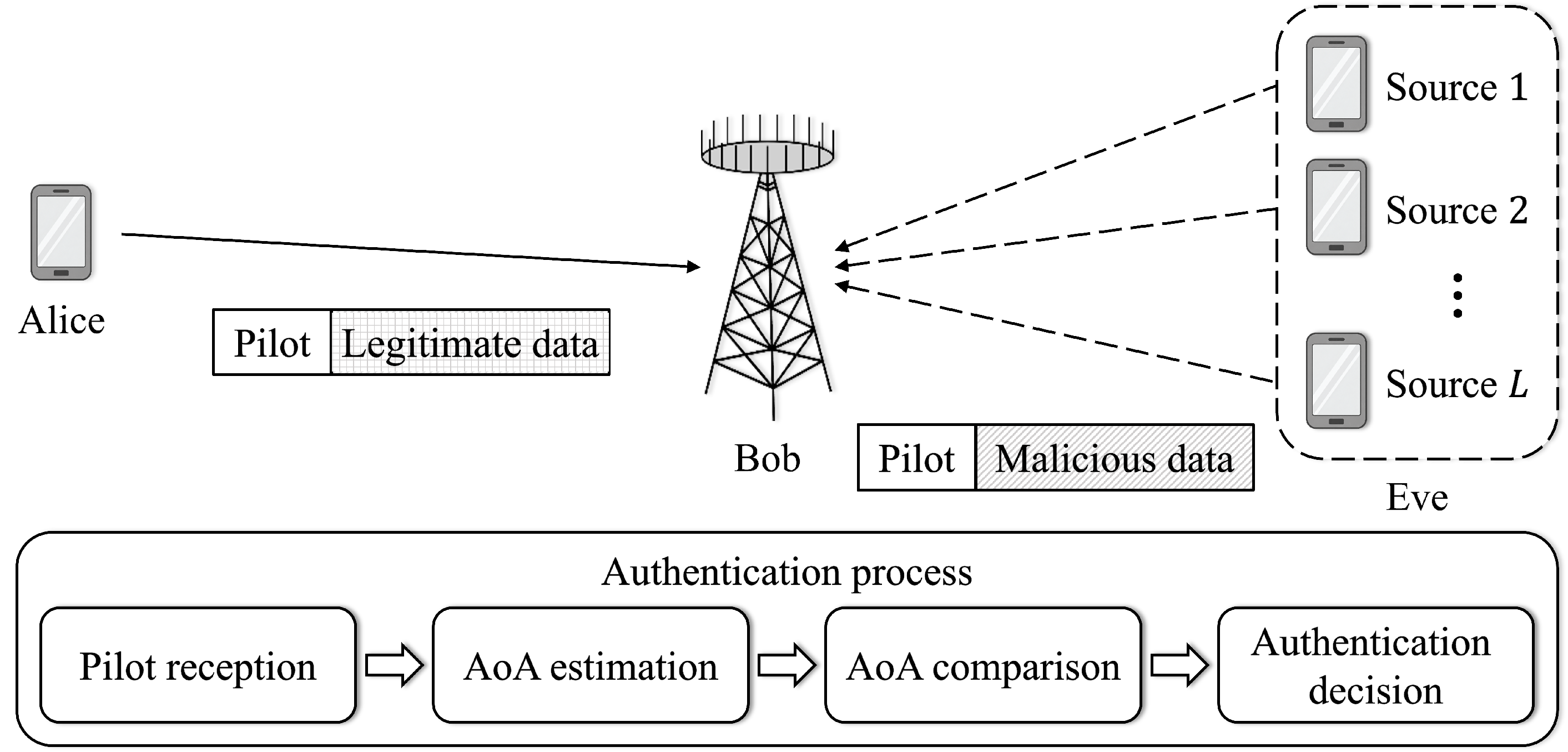, width=11.9cm}} 
    \caption{System model of AoA-based authentication under an impersonation attack.}
    \label{fig:System model}
\end{figure}

In this work, we consider an AoA-based authentication scenario under impersonation attacks, as illustrated in Fig. \ref{fig:System model}. A single-antenna legitimate transmitter (Alice) communicates with a multi-antenna receiver (Bob), while an attacker (Eve) attempts to forge Alice's identity by controlling $L$ distributed sources. Alice transmits a pilot sequence for channel estimation before data transmission. We consider a protocol in which the pilot sequence is generated deterministically according to a publicly specified rule and publicly available parameters. Eve can therefore reconstruct Alice's pilot sequence without access to upper-layer secret keys and transmit the same sequence in an attempt to impersonate Alice \cite{tugnait2018pilot}. Bob estimates the AoA from the received pilot sequence by means of a digital ULA. Bob then compares the estimated angle with Alice's AoA to distinguish Alice from Eve. We note that Bob has prior knowledge of Alice's AoA through upper-layer protocols that guarantee its authenticity. In this setup, Eve attempts to deceive Bob by transmitting precoded signals using $L$ distributed sources. We also assume that Eve operates at the same carrier frequency as Alice, making it impossible for Bob to distinguish between Eve and Alice based solely on their carrier frequencies.

\subsection{Signal Model}\label{Sec:Signal Model}
Under the assumption of a narrowband far-field propagation model, the signals transmitted by Alice and Eve arrive at Bob as plane waves. Then, the steering vector associated with Alice's AoA $\theta$ is given by
\begin{align}
\mathbf{a} =[1,e^{-j\kappa\sin\theta},\cdots,e^{-j\kappa(M-1)\sin\theta}]^{\top},
\end{align}
with $\kappa = 2\pi d/\lambda$, where $M$ and $d$ represent the number of antennas and the antenna spacing of Bob's ULA, respectively, and $\lambda$ denotes the carrier wavelength. During the pilot transmission, the received signal at Bob from Alice is given by
\begin{align}
\mathbf {y}[n]=\mathbf {a}s[n]+\mathbf {z}[n],~~ n\in\mathcal{N},
\end{align}
where $s[n]$ is the $n$-th symbol of the pilot sequence, $\mathbf{z}[n]\sim\mathcal{CN}(\mathbf{0}_M,\sigma^2 \mathbf{I}_M)$ is the additive white Gaussian noise (AWGN) vector for the Alice-Bob link, and $\mathcal{N} = \{0,1,\cdots,N-1\}$ is the set of pilot symbol indices, with $N$ being the pilot sequence length. 

Similarly, the received signal from the $L$ distributed sources controlled by Eve is given by
\begin{align}
\hat {\mathbf {y}}[n]=\hat {\mathbf {A}}\hat {\mathbf w}s[n]+\hat {\mathbf {z}}[n],~~ n\in\mathcal{N},
\end{align}
where $\hat{\mathbf{A}} =[\hat{\mathbf{a}}_0,\hat{\mathbf{a}}_1,\cdots,\hat{\mathbf{a}}_{L-1}]\in\mathbb{C}^{M \times L}$ denotes the steering matrix of Eve's multiple sources, $\hat {\mathbf w} \in \mathbb{C}^L$ is the precoding vector, and $\hat {\mathbf {z}}[n]\sim\mathcal{CN}(\mathbf{0}_M,\hat{\sigma}^2 \mathbf{I}_M)$ is the AWGN vector for the Eve-Bob link. Here, $\hat{\mathbf{a}}_l \in \mathbb{C}^M$ is the steering vector for Eve's $l$-th source, which can be expressed as 
\begin{align}
\hat{\mathbf{a}}_l =[1,e^{-j\kappa\sin\hat{\theta}_l},\cdots,e^{-j\kappa(M-1)\sin\hat{\theta}_l}]^{\top},~~ l\in\mathcal{L},
\end{align}
where $\hat{\theta}_l$ is the AoA of Eve's $l$-th source and $\mathcal{L}=\{0,1,\cdots,L-1\}$. Eve designs precoding across its distributed sources to bypass Bob's authentication and inject malicious data into the legitimate communication link.
In this study, the legitimate and adversarial links are each assumed to have a single line-of-sight (LoS) path. In practical environments, multipath propagation may provide additional opportunities for impersonation attacks \cite{pham2026leveraging}. However, to investigate whether Eve can bypass AoA-based authentication even with fewer propagation paths to exploit, we focus on LoS-based attacks and leave the analysis of multipath scenarios as future work.

\subsection{AoA-Based Authentication}\label{Sec:AoA-Based Authentication}
For authentication, Bob needs to estimate the AoA from the received signals, which are transmitted by either Alice or Eve. Various efficient methods have been developed to extract directional information from received signals, including fast Fourier transform (FFT)-based, Capon, and MUSIC methods. The FFT-based method suffers from limited angular resolution \cite{harter2011three}, and the Capon method shows poor estimation accuracy under highly noisy conditions \cite{weber2009analysis}. For this reason, we assume that Bob employs the MUSIC method to achieve high-resolution AoA estimation by searching for steering vectors orthogonal to the noise subspace \cite{schmidt1986multiple}.

Based on the estimated AoA, Bob determines whether the received signal originates from Alice or not by comparing it with Alice's known AoA. Accordingly, the authentication process is formulated as a binary hypothesis test:
\begin{align}
&\mathcal{H}_0: \text{The signal is transmitted by Alice}, \\
&\mathcal{H}_1: \text{The signal is not transmitted by Alice}.
\end{align}
If the difference between the two angles is below a predefined threshold $\tau$, Bob accepts the received signal as legitimate; otherwise, it is classified as an impersonation attempt. The decision rule is thus given by
\begin{align}
\bar{\mathcal{H}} = 
\begin{cases}
\mathcal{H}_0, ~~|\theta - \bar{\theta}| < \tau, \\
\mathcal{H}_1, ~~|\theta - \bar{\theta}| \geq \tau,
\end{cases}
\end{align}
where $\bar{\theta}$ is the AoA estimate obtained by Bob.

\section{Authentication Analysis}\label{Sec:Authentication Analysis}
In this section, we derive the optimal precoding at Eve to mimic the legitimate signal. After that, we provide a theoretical analysis that characterizes the conditions under which Eve can successfully impersonate Alice.

\subsection{Optimal Precoding for Impersonation Attacks}\label{Sec:Optimal Precoding for Impersonation Attacks}
The MSE between the received signals from Alice and Eve can be obtained as
\begin{align}
\zeta & =\frac{1}{N}\sum_{n \in \mathcal{N}}\mathbb {E}\Big[\|\mathbf {y}[n]-\hat {\mathbf {y}}[n]\|^{2}\Big] \nonumber \\
& = \frac{1}{N}\sum_{n \in \mathcal{N}}\mathbb {E}\Big[\| (\mathbf{a}-\hat{\mathbf{A}}\hat{\mathbf{w}})s[n] + \mathbf {z}[n] - \hat {\mathbf {z}}[n] \|^{2}\Big] \nonumber \\
& = \frac{1}{N}\sum_{n \in \mathcal{N}} \left( |s[n]|^2 \|\mathbf{a}\!-\!\hat{\mathbf{A}}\hat{\mathbf{w}}\|^2  \!+\! \mathbb {E}\big[\|\mathbf {z}[n] \|^{2}\big] \!+\!  \mathbb {E}\big[\|\hat {\mathbf {z}}[n] \|^{2}\big] \right)  \nonumber \\ 
 & = P\|\mathbf{a}-\hat{\mathbf{A}}\hat{\mathbf{w}}\|^2 + M\sigma^2 + M\hat{\sigma}^2,
 \label{eq:MSE}
\end{align}
where $P=\sum_{n \in \mathcal{N}} |s[n]|^2/N$ denotes the average power of the pilot sequence.
The MSE between the two signals serves as a measure of how closely Eve can mimic Alice's signal. Based on this criterion, the optimal precoding vector for the adversary can be determined by solving the MSE minimization problem defined as
\begin{align}
\min_{\hat{\mathbf{w}}\in\mathbb{C}^L}~\|\mathbf{a}-\hat{\mathbf{A}}\hat{\mathbf{w}}\|^2.
\end{align}
Following \cite{boyd2004convex}, this optimization problem is an unconstrained least-squares problem, whose minimum-norm solution can be obtained as
\begin{align}
\hat {\mathbf w}^{\star} = \hat{\mathbf {A}}^{\dagger}\mathbf{a}.
\label{eq:optimal precoding}
\end{align}
The squared norm is lower bounded as $\|\mathbf{a}-\hat{\mathbf{A}}\hat{\mathbf{w}}^{\star}\|^2 \geq 0$, and equality holds if and only if $\mathbf{a} \in \mathcal{R}(\hat{\mathbf {A}})$. This implies that Eve can successfully impersonate Alice only when Alice's steering vector lies in the column space of Eve's steering matrix. To further characterize the conditions for successful impersonation, we analyze the relationship between the steering vectors of Alice and Eve by considering two cases: $M>L$ and $M \leq L$.

It should be noted that knowledge of Alice's AoA is needed for Eve to obtain the optimal precoding vector in \eqref{eq:optimal precoding}. Although such knowledge may not always be available in practice, we adopt a worst-case adversarial model in which this AoA is perfectly known to Eve \cite{pham2026leveraging}, \cite{hassani2026near}. This setting allows us to characterize the spatial constraints on signal reproduction even with this information available to the adversary.

\subsection{Theoretical Analysis for $M>L$}\label{Sec:Theoretical Analysis1}
We first investigate the feasibility of successful impersonation when the number of antennas at the authenticator is greater than the number of adversarial sources. The result is summarized in Theorem \ref{thm_1}:

\vspace{2mm}
\begin{thm}\label{thm_1}
Suppose that $M>L$. Let $\Delta$ be the minimum squared error, i.e.,  $\Delta = \|\mathbf{a}-\hat{\mathbf{A}}\hat{\mathbf{w}}^{\star}\|^2$. Then, $\Delta = 0$ can be achieved if and only if  
\begin{align}
\sin\theta = \sin\hat{\theta}_l + k\lambda/d,
\label{eq:thm1_1}
\end{align}
for at least one $l\in\mathcal{L}$ and some $k\in\mathbb{Z}$.
\end{thm}
\begin{IEEEproof}
First, we assume that there exists at least one $l \in \mathcal{L}$ such that $e^{-j\kappa\sin\theta}=e^{-j\kappa\sin\hat{\theta}_l}$. By the periodicity of the complex exponential, this condition is equivalent to $\sin\theta = \sin\hat{\theta}_l + k\lambda/d$ for some $k\in\mathbb{Z}$. Furthermore, this implies that $\mathbf{a}=\hat{\mathbf{a}}_l$, and thus $\mathbf{a}$ can be expressed as a linear combination of the columns of $\hat{\mathbf {A}}$. Consequently, Eve can perfectly reproduce the legitimate signal under optimal precoding, which guarantees $\Delta = 0$.

Conversely, suppose that $e^{-j\kappa\sin\theta} \neq e^{-j\kappa\sin\hat{\theta}_l}$ for all $l \in \mathcal{L}$. Under the assumption that $M>L$, we have $\operatorname{rank}(\hat{\mathbf {A}})\leq L<M$. Then, the augmented matrix $[\mathbf{a},\hat{\mathbf {A}}]$ is introduced to examine the linear dependence between $\mathbf{a}$ and the columns of $\hat{\mathbf {A}}$. Note that this augmented matrix has a Vandermonde structure. It follows that the condition $e^{-j\kappa\sin\theta} \neq e^{-j\kappa\sin\hat{\theta}_l}$ for all $l$, together with $\operatorname{rank}(\hat{\mathbf {A}})\leq L<M$, implies $\operatorname{rank}([\mathbf{a},\hat{\mathbf {A}}])=\operatorname{rank}(\hat{\mathbf {A}})+1$ by the rank property of Vandermonde matrices \cite{shan1985spatial}. This rank relationship demonstrates that $\mathbf{a}$ is linearly independent of any basis for $\mathcal{R}(\hat{\mathbf {A}})$. Consequently, the minimum squared error is strictly positive, i.e., $\Delta>0$. Combining these results, we establish the necessary and sufficient condition for $\Delta=0$, which completes the proof. 
\end{IEEEproof}
\vspace{2mm}

Theorem \ref{thm_1} reveals that, when $M>L$, the feasibility of an impersonation attack depends on the relative geometry of Alice, Eve, and Bob. Specifically, Eve cannot achieve $\Delta=0$ unless at least one of its sources satisfies the AoA condition, regardless of the precoding strategy.

\subsection{Theoretical Analysis for $M\leq L$}\label{Sec:Theoretical Analysis2}
We now derive the conditions for successful impersonation attacks when the number of adversarial sources is greater than or equal to the number of receive antennas. The result is summarized in the following theorem:

\vspace{2mm}
\begin{thm}\label{thm_2}
When $M\leq L$, the minimum squared error satisfies $\Delta = 0$ if and only if either
\begin{align}
\sin\theta = \sin\hat{\theta}_l + k\lambda/d,
\label{eq:thm2_1}
\end{align}
for at least one $l\in\mathcal{L}$ and some $k\in\mathbb{Z}$, or there exist $l_1,\cdots,l_M \in \mathcal{L}$ such that
\begin{align}
\sin\hat{\theta}_{l_i} \neq \sin\hat{\theta}_{l_j} +k\lambda/d,
\label{eq:thm2_2}
\end{align}
for all $i,j\in\{1,\cdots,M\}$ with $i \neq j$ and all $k\in \mathbb{Z}$.
\end{thm}
\begin{IEEEproof}
The sufficiency of the condition in \eqref{eq:thm2_1} follows from the same logic as in the proof of Theorem \ref{thm_1}. We now consider the second condition and suppose that there exist $l_1,\cdots,l_M$ satisfying \eqref{eq:thm2_2}. This condition can be written as $e^{-j\kappa\sin\hat{\theta}_{l_i}} \neq e^{-j\kappa\sin\hat{\theta}_{l_j}}$ for all $i\neq j$. From the Vandermonde structure of $\hat{\mathbf{A}}$, the steering vectors $\hat{\mathbf{a}}_{l_1},\cdots,\hat{\mathbf{a}}_{l_M}$ are linearly independent. As a result, the steering matrix $\hat{\mathbf{A}}$ has full row rank, which guarantees that any vector in $\mathbb{C}^M$, such as Alice's steering vector $\mathbf{a}$, can be expressed as a linear combination of its columns.

To establish the necessity, assume that neither of the two conditions holds. 
Under this assumption, we obtain $\operatorname{rank}(\hat{\mathbf {A}})<M$ and $e^{-j\kappa\sin\theta} \neq e^{-j\kappa\sin\hat{\theta}_l}$ for all $l\in\mathcal{L}$. The augmented matrix $[\mathbf{a},\hat{\mathbf {A}}]$ therefore has a rank that is one higher than that of $\hat{\mathbf {A}}$ due to its Vandermonde structure. In other words, the minimum achievable squared error is nonzero even with optimal precoding. The proof is completed by combining the sufficiency and necessity arguments.
\end{IEEEproof}
\vspace{2mm}

One important observation from Theorem \ref{thm_2} is that perfect signal reproduction is possible even when none of Eve's steering vectors is identical to Alice's steering vector, provided that sufficient angular diversity is available. 
This suggests that a larger number of adversarial sources provides Eve with more opportunities to successfully impersonate the legitimate user.

The results of Theorems \ref{thm_1} and \ref{thm_2} can be further simplified under specific antenna spacing and AoA conditions, as stated in the following corollary.

\vspace{2mm}
\begin{cor}\label{cor_1}
Let $d\leq\lambda/2$ and $\theta,\hat{\theta}_l\in(-\pi/2,\pi/2)$ for all $l\in\mathcal{L}$. Then, the necessary and sufficient condition for $\Delta=0$ is that Eve has either at least one source with the same AoA as Alice or at least $M$ sources with mutually distinct AoAs.
\end{cor}
\begin{IEEEproof}
Under the given conditions, we have $\lambda/d\geq 2$ and $|\sin\theta_1-\sin\theta_2|<2$ for any two AoAs $\theta_1,\theta_2\in(-\pi/2,\pi/2)$. Thus, the equalities in \eqref{eq:thm1_1} and \eqref{eq:thm2_1} can hold only for $k=0$, while the inequality in \eqref{eq:thm2_2} is satisfied for all nonzero $k$. 
Since the sine function is strictly increasing over $(-\pi/2,\pi/2)$, the conditions in \eqref{eq:thm1_1} and \eqref{eq:thm2_1} reduce to $\theta=\hat{\theta}_l$ for some $l\in\mathcal{L}$. Similarly, \eqref{eq:thm2_2} can be simplified to $\hat{\theta}_{l_i}\neq\hat{\theta}_{l_j}$ for all $i,j\in\{1,\cdots,M\}$ with $i\neq j$.
\end{IEEEproof}
\vspace{2mm}

Corollary \ref{cor_1} shows that, without spatial aliasing, the feasibility of perfect signal reproduction can be characterized by the AoAs of Alice and Eve's sources along with the number of receive antennas. This provides a simplified interpretation of the geometric conditions for successful impersonation attacks.
We also note that the theoretical analysis in Sec. \ref{Sec:Authentication Analysis} can be directly extended even when Alice and Eve operate at different carrier frequencies and/or Bob is equipped with a uniform planar array.

\section{Numerical Results}\label{Sec:Numerical Results} 
In this section, we present numerical results to validate the theoretical analysis and evaluate the security performance of AoA-based authentication under impersonation attacks. In the simulations, the carrier frequency is set to $2.8 \textrm{ GHz}$, and the authenticator is equipped with a half-wavelength spaced ULA. For pilot transmission, we consider a Zadoff–Chu sequence \cite{chu1972polyphase} of length $N=20$ with a root index of $u=3$, whose $n$-th symbol is given by
\begin{align}
s[n]= \sqrt{P} e^{j u \pi n^2/N},~~ n\in\mathcal{N}.
\end{align}
Under this setup, we investigate two cases: \lowercase\expandafter{\romannumeral1}) \textit{condition in \eqref{eq:thm1_1} or \eqref{eq:thm2_1} holds}, where exactly one of Eve's sources has the same AoA as Alice; \lowercase\expandafter{\romannumeral2}) $\textit{condition in \eqref{eq:thm2_2} holds}$, where all of Eve's sources have AoAs different from Alice's AoA and from each other. Unless otherwise specified, the AoAs of Alice and Eve's sources are generated from $\mathcal{U}(0,2\pi)$, with one of Eve's AoAs being identical to that of Alice in the first case. We consider the MSE in \eqref{eq:MSE} and the attack success probability as performance metrics. The attack success probability is defined as $\mathbb{P}[|\theta - \bar{\theta}| < \tau \,|\, \mathcal{H}_1]$. All simulation results are averaged over $10^6$ Monte Carlo trials.

\begin{figure}[t]
\centering 
    {\epsfig{file=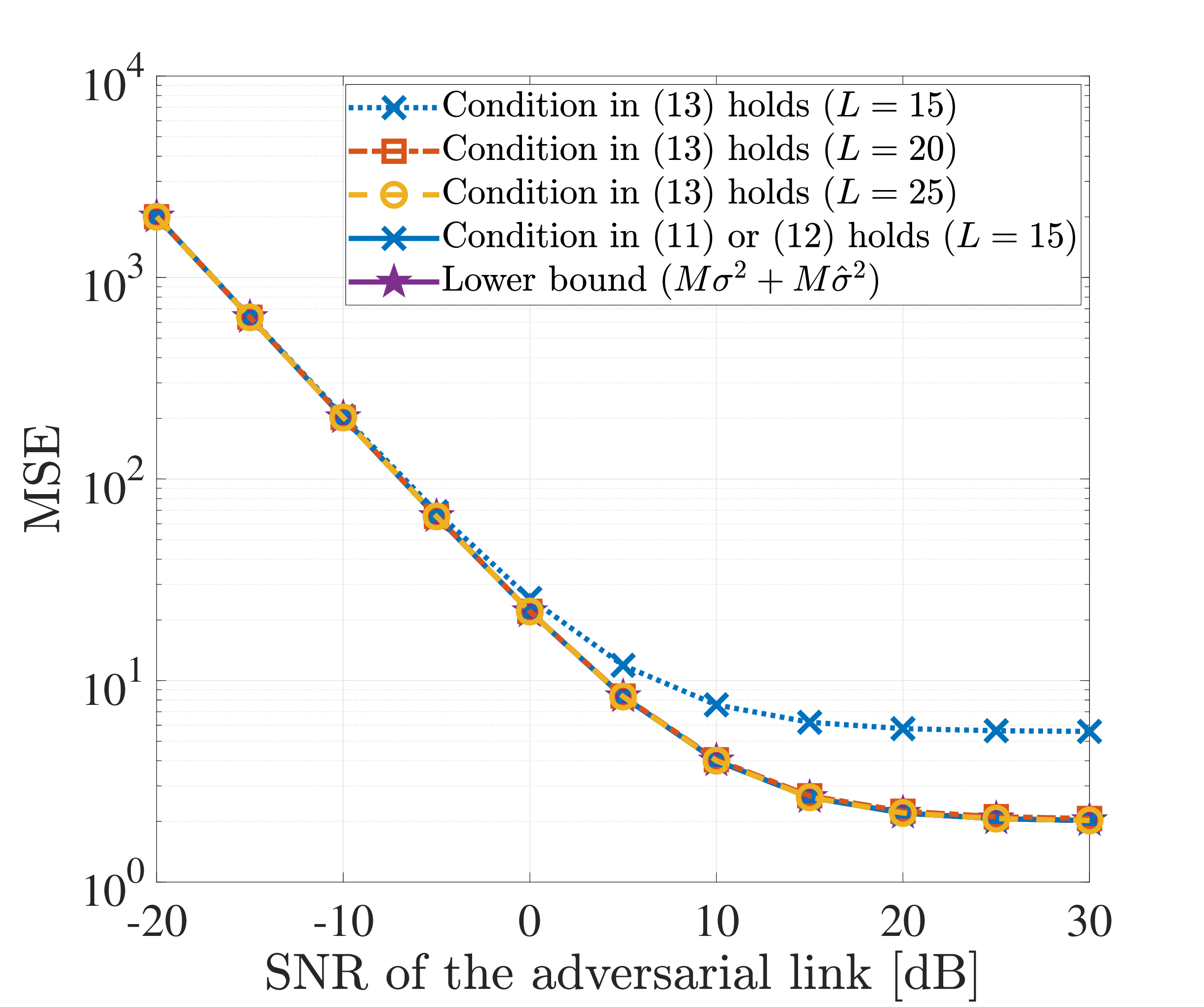, width=8.5cm}} 
    \caption{MSE versus SNR of the adversarial link ($M=20$ and $P/\sigma^2=10 \textrm{ dB}$).}
    \label{fig:MSE_vs_SNR_Result}
\end{figure}

Fig. \ref{fig:MSE_vs_SNR_Result} plots the MSE versus the signal-to-noise ratio (SNR) of the adversarial link. We first observe that higher adversarial SNRs make it more difficult for Bob to distinguish Eve from Alice due to the lower MSE between their signals at the authenticator. When \eqref{eq:thm2_2} holds, the MSE approaches the noise floor, i.e., $M\sigma^2+M\hat{\sigma}^2$, as the number of Eve's distributed sources increases and achieves it when $L=M$. In contrast, when \eqref{eq:thm1_1} or \eqref{eq:thm2_1} holds, the MSE can be reduced to the noise floor even when Eve has fewer distributed sources than the number of receive antennas at Bob.

\begin{figure}[t]
\centering 
    {\epsfig{file=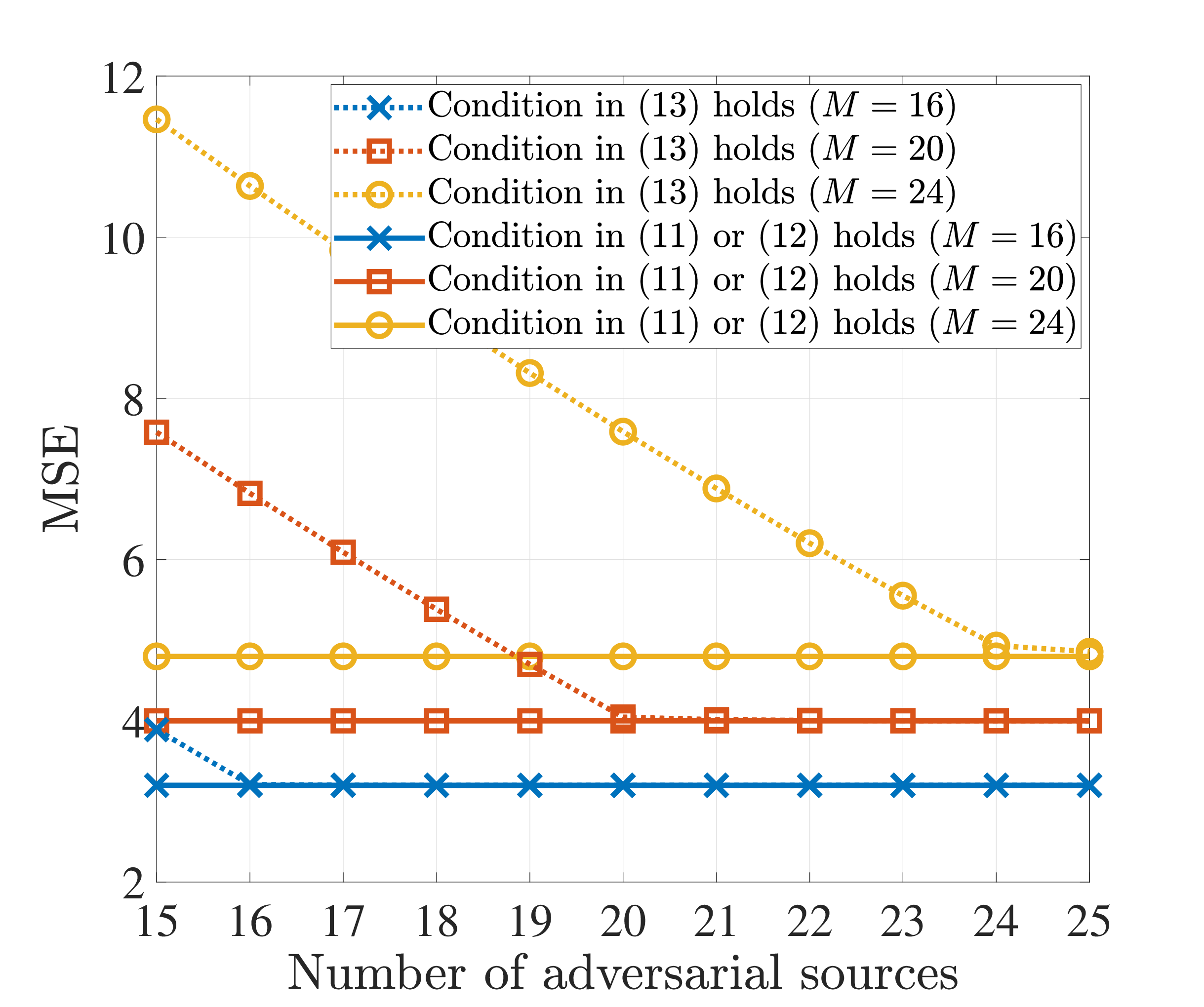, width=8.5cm}} 
    \caption{MSE versus the number of adversarial sources ($P/\sigma^2=10 \textrm{ dB}$ and $P/\hat{\sigma}^2=10 \textrm{ dB}$).}
    \label{fig:MSE_vs_L_Result}
\end{figure}

Fig. \ref{fig:MSE_vs_L_Result} shows the MSE versus the number of distributed sources controlled by Eve. The minimum MSE can be achieved regardless of the number of distributed sources when at least one of the sources lies in the same direction as Alice relative to Bob. Otherwise, the feasibility of successful impersonation is influenced by how many sources Eve controls compared with the number of Bob's receive antennas. These results provide verification of our theoretical analysis of AoA-based authentication.

\begin{figure}[t]
\centering 
    {\epsfig{file=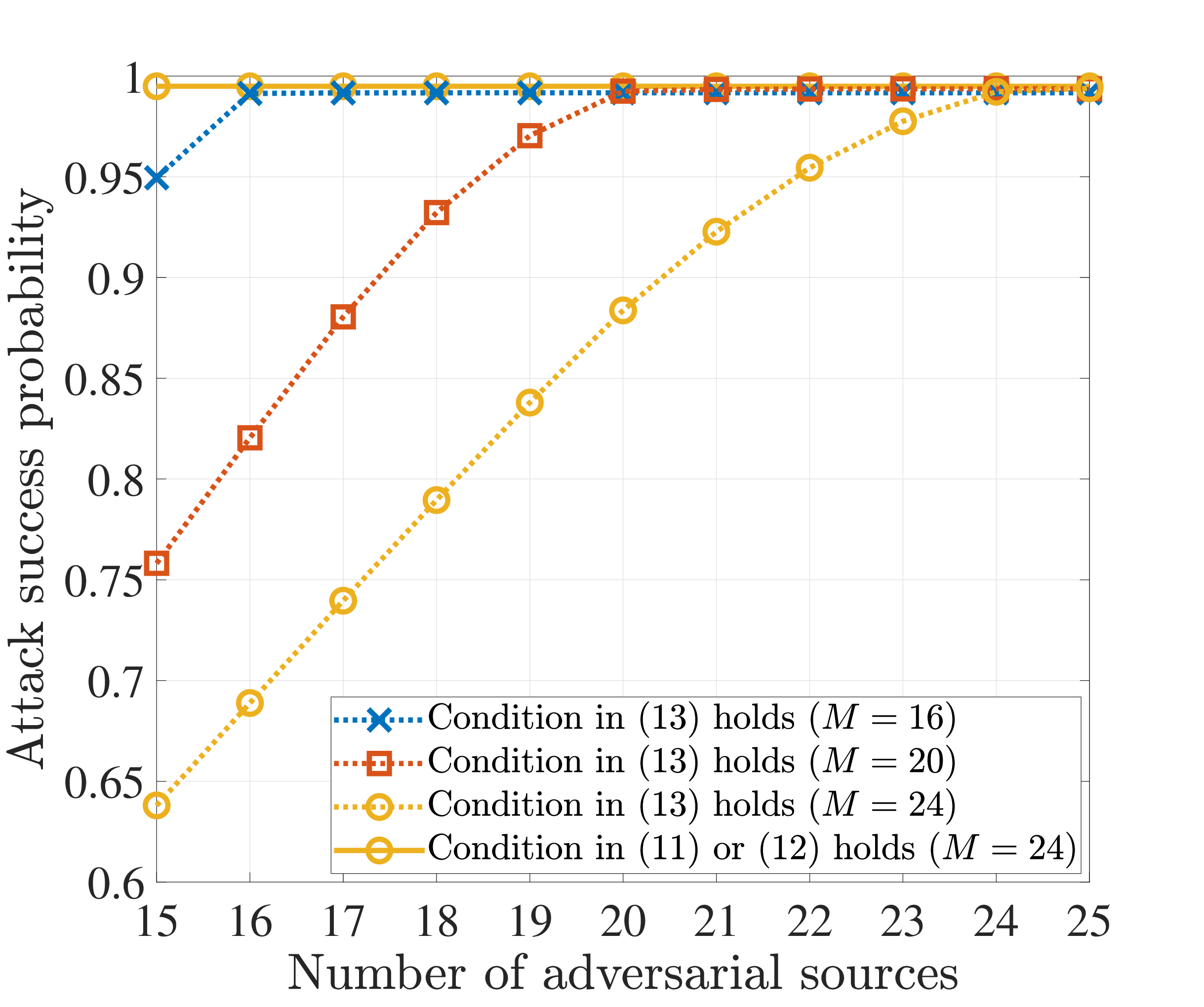, width=8.5cm}} 
    \caption{Attack success probability versus the number of adversarial sources ($P/\hat{\sigma}^2=20 \textrm{ dB}$).}
    \label{fig:MUSIC_Attack_Probability_vs_L_Result}
\end{figure}

Fig. \ref{fig:MUSIC_Attack_Probability_vs_L_Result} plots the probability of successful impersonation versus the number of adversarial sources. The AoAs of Alice and Eve's sources follow $\mathcal{U}(-\pi/2,\pi/2)$ to avoid angular ambiguity in the ULA. For AoA estimation, Bob performs a MUSIC search over $\bar{\theta} \in [-\pi/2,\pi/2]$ with a step size of $\pi/1800$. The authentication threshold is set to $\tau=\pi/180$. The figure shows that, even with a large number of receive antennas at Bob, AoA-based authentication cannot guarantee security if any of Eve's sources has an AoA identical to that of Alice. When none of Eve's sources has the same AoA as Alice, authentication is more secure if Bob is equipped with sufficiently many receive antennas.

\section{Conclusion}\label{Sec:Conclusion}
In this letter, we investigated cooperative impersonation attacks against AoA-based authentication in a digital ULA system. We derived the optimal precoding vector for multiple spatially distributed adversarial sources and established necessary and sufficient conditions under which Eve can forge the legitimate transmitter's spatial signature at the authenticator. Our theoretical analysis revealed that when the number of adversarial sources is smaller than that of receive antennas, Eve can perfectly forge the legitimate transmitter's spatial signature only if a specific AoA condition is satisfied by at least one adversarial source. Otherwise, AoA-based authentication may be vulnerable to cooperative impersonation attacks under sufficient angular diversity among adversarial sources, regardless of whether an individual source satisfies this AoA condition. Via simulations, we validated the theoretical findings and evaluated the security performance of AoA-based authentication. As part of our future work, we plan to delve into the feasibility and performance of cooperative impersonation attacks in multipath propagation environments under practical transmit power constraints.

\bibliographystyle{IEEEtran}
\bibliography{ref}

\end{document}